\documentclass[aps,prl,twocolumn,groupedaddress,showpacs]{revtex4}

\begin{document}

\title{
Comment on ``Using Three-Body Recombination to Extract \\
Electron Temperatures of Ultracold Plasmas''
}

\author{Yurii V. Dumin}

\email[Electronic address: ]{dumin@yahoo.com}

\affiliation{
Max Planck Institute for the Physics of Complex Systems,
Noethnitzer Strasse 38, 01187 Dresden, Germany
}

\altaffiliation[On leave from: ]{
IZMIRAN, Russian Academy of Sciences,
Troitsk, Moscow reg., 142190 Russia
}

\date{
December 30, 2008
}

%

\maketitle

In the recent work~\cite{fle07} Fletcher \textit{et al.}\ reported on
the novel experimental technique, enabling to measure the temperature of
the expanding ultracold plasmas over a considerable time interval
(up to $ 60{\div}70$~$\mu$s). It was unexpectedly found that the electron
temperature dropped with time as $ T_e (t) \sim t^{- \alpha} $
with $ \alpha = 1.2 \pm 0.1 \approx 1 $ instead of $ \alpha = 2 $,
which would be expected for the adiabatic cooling of electrons
in the cloud expanding linearly in time.
The above-cited authors supposed that `the difference is likely due to
the significant heating effects from 3-body recombination'
(\textit{i.e.}\ the inelastic processes), but they did not provide
sufficient quantitative estimates supporting such a conclusion.
The aim of the present comment is to mention that the experimentally
revealed $ t^{-1} $-dependence can be explained under quite general
assumptions by the purely elastic processes in the ultracold plasma,
as it was done a few years ago in our work~\cite{dum00}.
Briefly speaking, the proof of the universality of
the $ t^{-1} $-behavior consists of the three main steps.

Firstly, we start from the most general form of the electron
distribution function:
\begin{eqnarray}
&&
f( \textbf{r}_{e 1}, \dots , \textbf{r}_{e N_e},
\textbf{v}_{e 1}, \dots , \textbf{v}_{e N_e} ) =
A_f \exp \! \Big\{
  \!\! - \! \frac{1}{k_{\rm B} T_e}
\nonumber \\
&& \quad
  \times \Big[
  \, \sum_{n} \, \frac{ m_e \textbf{v}_{en}^2 }{ 2 }
  + \, U( \textbf{r}_{e 1}, \dots , \textbf{r}_{e N_e},
  \textbf{r}_{i 1}, \dots , \textbf{r}_{i N_i} )
\Big] \! \Big\} \, ,
\nonumber
\end{eqnarray}
where
$ \textbf{r}_{en} $ and $ \textbf{v}_{en} $ are the coordinates and
velocities of electrons,
and $ \textbf{r}_{in} $ are the coordinates of ions
in a sufficiently small volume of plasma, moving with
a definite macroscopic velocity.
(The kinetic energy of ions could be also written here,
but it is typically much less than the electronic one.)
Despite a very complex form of the potential energy~$U$
in the regime of strong coupling,
calculation of the thermodynamic quantities depending only
on the velocities is quite easy,
because the integrals of~$U$ in the numerator and denominator cancel
each other.
Particularly, the average kinetic energy per one particle
turns out to be
$ \langle k \rangle = (3/2) \, k_{\rm B} T_e $.
This looks formally as an expression for the ideal gas,
but it is actually applicable to the plasma with any strength of
the Coulomb's interaction~$U$ between the particles,
including the state of deep cooling.

Secondly, the average kinetic energy~$ \langle k \rangle $
can be related to the average potential energy~$ \langle u \rangle $
by the well-known virial relation for the Coulomb's field:
$ \langle k \rangle = (1/2) \: | \langle u \rangle | $,
which is also valid at the arbitrary intensity of
interparticle interactions.
Strictly speaking, the virial theorem takes place only for
the systems experiencing a finite (\textit{i.e.}\ restricted
in space) motion. Nevertheless, it should be approximately
applicable also to the small (but macroscopic) volume elements
of the ultracold plasma cloud due to attractive potential
by the ions, resulting in the quasi-confined electron
motion~\cite{fle06}.

Thirdly, the average potential energy can be evidently expressed
in terms of the average interparticle distance:
$ \langle u \rangle \sim \, e^2 \! / \langle r \rangle \sim \,
e^2 n^{1/3} $, where $e$~is the electron charge,
and $n$~is the concentration of charged particles.

Finally, combining the above-written formulas, we get
$ T_e \! \sim n^{1/3} $. In particular, if the cloud expands
linearly in time (and, consequently, its concentration changes
as~$ t^{-3} $), then $ T_e \! \sim t^{-1} $.
Such kind of the temporal dependence was initially derived
in~\cite{dum00} for a uniform cloud whose boundary moved by
a linear law, but it should be approximately the same
for a cloud with Gaussian density distribution
whose dispersion (\textit{i.e.}\ the characteristic size)
increases almost linearly at large times
(see formula~(1) and below in paper~\cite{fle07}).

Therefore, $t^{-1}$-dependence revealed in the above-cited
experiment should be primarily a manifestation of ``virialization''
of the electron velocities in the regime of strong electron--ion
correlations, while the heat release by inelastic processes
should be of secondary importance (it might be responsible,
particularly, for changing the exponent from $-1$ to $-1.2$).
This is confirmed also by our \textit{ab initio}
molecular-dynamic simulations taking into account the strong
electron--ion correlations (\textit{i.e.} not based on the
PIC method, Vlasov approximation for electrons, \textit{etc}.,
which are applicable only to small-angular scattering)%
\footnote{
See EPAPS Document No.\ E-PRLTAO-XXX-XXXXXX.
}.

\end{document}